\documentclass[english]{llncs}
\usepackage{amsmath,amssymb,verbatim,babel,wrapfig}
\usepackage{graphicx}

\def\eps{\varepsilon}

\def\R{\mathbb{R}}
\def\FD{Fr\'{e}chet distance}
\def\WFD{weak Fr\'{e}chet distance}

\def\SCEP{simple curve embedding problem}

\def\SOrder{stringing order}
\def\SCurve{stringing curve}
\def\SCurves{stringing curves}
\def\SigC{signal curve}
\def\SigCs{signal curves}
\def\HCurve{hole curve}
\def\HCurves{hole curves}
\usepackage{xspace}

\def\HD{Hausdorff distance}
\def\curve{f}
\def\surface{S}
\def\mappedCurve{f'}
\def\problem{simple curve embedding problem}

\begin{document}

\setcounter{page}{1}
\pagestyle{plain}

\title{Simple Curve Embedding\thanks{This work has been supported by National
      Science Foundation grant CCF-0643597.}}

\author{Jessica Sherette\inst{1} \and Carola Wenk\inst{2}}

\institute{Department of Computer Science, University of Texas at San Antonio, USA, \email{jsherett@cs.utsa.edu}
\and 
Department of Computer Science, Tulane University, USA, \email{cwenk@tulane.edu}}

\maketitle

\begin{abstract}
Given a curve $\curve$ and a surface $\surface$, how hard is it to find a simple curve $\mappedCurve\subseteq\surface$ that is the most similar to $\curve$? 

We introduce and study this {\bf \problem} for piecewise linear curves and surfaces in $\R^2$ and $\R^3$, under \HD, \WFD, and \FD\ as similarity measures for curves. Surprisingly, while several variants of the problem turn out to have polynomial-time solutions, we show that in $\R^3$ the \problem\ is NP-hard under \FD\ even if $\surface$ is a plane, as well as under \WFD\ if $\surface$ is a terrain. Additionally, these results give insight into the difficulty of computing the \FD\ between surfaces, and they imply that the partial \FD\ between non-planar surfaces is NP-hard as well.

\end{abstract}
%
%

\section{Introduction}
Given a curve $\curve$ and a surface $\surface$, how hard is it to find a simple curve $\mappedCurve\subseteq\surface$ that is the most similar to $\curve$? 
We study this {\bf \problem} for piecewise linear curves and surfaces in $\R^2$ and $\R^3$, under \HD, \WFD, and \FD\ as similarity measures. Surprisingly, while several variants of the problem turn out to have polynomial-time solutions, we show that in $\R^3$ the \problem\ is NP-hard under \FD\ even if $\surface$ is a plane, as well as under \WFD\ if $\surface$ is a terrain. It then follows that the partial \FD\ between non-planar surfaces is NP-hard as well.

Curve embedding problems have received attention in the literature lately, most prominently in the case of map-matching in road networks \cite{aerw-mpm-03,bpsw-mmvtd-05,nk-hmmm-09,wrp-abnns-06}. Here, given a trajectory of location measurements, the task is to find a, not necessarily simple, path in a road network graph that is most similar to the trajectory. 
 There are polynomial-time map-matching algorithms for the \FD\ \cite{aerw-mpm-03,bpsw-mmvtd-05} and for the \WFD\ \cite{bpsw-mmvtd-05}. Har-Peled and Raichel \cite{hr-fdre-11} generalized the problem to finding two, not necessarily simple, curves in each of two given input simplicial complexes.

The importance of simplicity on the other hand has been recognized in the area of morphing, where a continuous mapping is sought that continuously deforms one shape into another. If the two shapes are simple curves, then such a morphing (homotopy) should ideally enforce that every intermediate curve is also simple. Even though the \FD\ is based on computing continuous assignments between points on the two curves, linear interpolations along these assignments can yield morphings in which intermediate curves self-intersect. Efrat et al.\cite{Efrat2002} showed that a geodesic variant of the \FD\ yields morphings with simple intermediate curves and can be computed in polynomial time.
This exploits the fact that shortest paths in a polygon do not cross. Chambers and Letscher \cite{cl-ifd-11} introduced the notion of the isotopic \FD\ which requires intermediate curves to be simple, however, there is no algorithm to compute it. 

Simple curve embedding is also closely related to computing the \FD\ between two surfaces. Computing the \FD\ between two surfaces is NP-hard \cite{bbs-fdsss-10,g-cmsbg-99} and upper-semi-computable \cite{ab-cwcsb-10}, but it is not known whether it is computable. 
Polynomial-time algorithms exist, however, to compute the \FD\ between two planar polygons \cite{bbw-cfdsp-06} and to approximate it between two folded polygons (piecewise linear surfaces with acyclic dual graphs) \cite{cdhsw-cfdfp-11}. Both algorithms are based on embedding straight-line diagonals from one surface into the other surface while avoiding any crossings of the embedded curves. In the folded polygons case, this leads to the necessity of ``untangling'' embedded curves that cross. Untangling a single self-intersecting curve is exactly the \problem. It is not known whether computing the \FD\ between folded polygons is NP-hard. Curve untangling has also recently been studied for the problem of minimizing the number of flip operations to convert a non-planar graph drawing of a planar graph into a planar one \cite{gkossw-upg-09}.
%
%
%
%

\noindent{\bf Our contributions.}
We consider $\curve$ to be piecewise linear and simple in $\R^3$, or self-intersecting in $\R^2$. 
We first sketch polynomial-time algorithms 
for possibly self-intersecting $\mappedCurve$ under \WFD\ and \FD.
%
We then prove that the \problem\ is NP-hard under \FD\ if $\surface$ is a plane with holes. We extend this proof to show NP-hardness under \FD\ if $\surface$ is a plane, and under \WFD\ if $\surface$ is a terrain. It then follows that the partial \FD\ for folded polygons is NP-hard, even if $P$ is a polygonal strip and $Q$ is the plane. 

\section{Preliminaries \label{Preliminaries}}
Given a curve $\curve$ and a surface $\surface$, the {\bf \problem} is to find a 
simple curve $\mappedCurve\subseteq\surface$ that has the smallest distance to $\curve$. 
Its {\em decision variant} is, given an additional parameter $\eps>0$, to decide whether there is a
simple curve $\mappedCurve\subseteq\surface$ that is in distance at most
$\varepsilon$ to $\curve$.
We call such a curve $\mappedCurve$ a {\em valid embedded curve}.

The {\em \FD} is defined for two 
hypersurfaces $P,Q:[0,1]^{k}\rightarrow\mathbb{R}^{d}$ as
\begin{align*}
    \delta_{F}(P,Q) = \inf_{\sigma:[0,1]^k\rightarrow [0,1]^k} \sup_{p\in [0,1]^k}
    \|P(p)-Q(\sigma(p))\|,
\end{align*}
where $\sigma$ ranges over orientation-preserving homeomorphisms, $k\leq d$, and $\|\cdot\|$ is a metric which in our case is the Euclidean norm. A common intuitive explanation of the \FD\ in the case of curves, for $k=1$, is as follows: Suppose a man walks along one curve, a dog walks along the other, and they are connected by a leash.  They can vary their relative speeds but cannot move backwards.  Then the \FD\ of the curves is the minimum leash length required for the man and dog to walk along these curves. In the {\em \WFD}, $\sigma$ ranges over surjective continuous maps instead.

For two coplanar triangulated simple polygons $P, Q$ and some $\varepsilon > 0$, the {\em partial \FD\ problem} is to decide whether there exists a simple polygon $R \subseteq Q$ such that
$\delta_{F}(P,R) \leq \varepsilon$.

\section{Polynomial-Time Algorithms \label{PolyTime}}

In this section we briefly examine some variants of the \problem\ which can be solved in polynomial time.

A natural first step is to drop the simplicity constraint and solve the curve embedding problem when the output curve $\mappedCurve$ might self-intersect. Under \WFD\ and for a triangulated surface $\surface$, the curve embedding problem can be solved in polynomial time. This follows directly from the work of Har-Peled and Raichel \cite{hr-fdre-11}.
For the \FD, the problem differs from \cite{hr-fdre-11} in that the curve $\curve$ has to be traversed in a monotone fashion. For a given $\eps$, consider the three-dimensional free space defined by the parameter spaces of $\curve$ and $\surface$. This is essentially a subset of the product space consisting of all point pairs that are in distance at most $\eps$. The problem reduces to finding a path in this free space that is monotone in $\curve$. Similar to the algorithm to compute the \FD\ between curves \cite{ag-cfdbt-95}, one can propagate reachability information along the boundaries of convex free space cells, filling a dynamic programming table ``from bottom to top'' in the direction of $\curve$. Propagation across one cell is easy: Only when propagating in a direction other than $\curve$, the reachable space needs to be cropped with a line in order to encode monotonicity constraints along $\curve$. The complexity of the reachable space remains constant on each cell boundary, and a possibly self-intersecting $\mappedCurve$ with $\delta_F(\curve,\mappedCurve)\leq\eps$ can be computed in $O(|\curve||\surface|)$ time.


While we show below that the \problem\ is NP-hard for many variants, interestingly the simple and unrestricted variants of the problem are equivalent for the \HD: For any embedded curve $\mappedCurve\subseteq\surface$ within \HD\ $\eps$ to $\curve$ one can generate a simple curve $f''$ which is within \HD\ $\eps$ to $\curve$ as follows: Treat $\mappedCurve$ as a graph with vertices added at each point where it crosses itself, and then traverse the graph with a simple curve $f''$, similar to the method outlined in Section \ref{Polyholes} for planar graphs.  By definition, this does not change the \HD.

Finally, let $\surface$ be a simple polygon and $\curve$ be a polygonal curve that is both simple and coplanar with $\surface$. We can extend the polynomial-time algorithm for computing the partial \FD\ for surfaces \cite{sw-pmsufd-12} by treating the curve $\curve$ as a very narrow simple polygon $P_f$.  The interior of $P_f$ can be made arbitrarily small and, thus, will have negligible impact on the output of the algorithm.  
%
In Section \ref{PartialFD-Proof}, we use a similar approach to obtain a hardness result for computing the partial \FD\ for surfaces.

\section{NP-hardness in the Plane with Holes \label{Polyholes}}

%

In this section we outline a polynomial-time reduction from Planar 3SAT to the \problem\ under \FD\ when the surface $\surface$ is a plane with holes and the curve $\curve$ is self-intersecting and lies entirely in the plane.
Let $F$ be a Boolean 3CNF formula. Its associated (bipartite) graph $G_F$ has one vertex for each variable and for each clause in $F$, and an edge connects variable and clause if that variable or its negation appears in that clause in $F$. An instance of Planar 3SAT is a Boolean 3CNF formula $F$ whose associated graph $G_F$ is planar. Planar 3SAT is NP-complete \cite{l-pftu-82}.
%
%
This section is devoted to proving the following theorem.

\begin{theorem}%
    \label{Polyholes-Thm}
  Let $\curve$ be a polygonal curve and $\surface$ be a plane with holes. The decision version of the simple curve embedding problem under \FD\ for $\curve$, $\surface$, and $\eps$ is NP-hard.
%
\end{theorem}

Assume we are given an instance $F$ of Planar 3SAT, and a straight-line embedding\footnote{Such an embedding can be computed in linear time.} of its associated planar graph $G_F$. Our reduction modifies this embedding by replacing vertices and edges with gadgets. 
Each edge in $G_F$ is replaced with a WIRE gadget, each variable-vertex with a VARIABLE gadgets, and each clause-vertex with a CLAUSE gadget.  If a variable appears as a negative instance in a clause then a NOT gadget is inserted along the WIRE connecting them in the construction.
Our reduction ensures that there exists a satisfying assignment of the variables if and only of there exists a valid embedded curve. 
%
%

\subsection{Stringing Order \label{Polyholes-StringingOrder}}
In our reduction, the gadgets are formed using holes in the surface $\surface$ and pieces of the curve $\curve$. We refer to these pieces as {\em \SigCs}; these are drawn in red, blue, and green in the figures, and possible embeddings for them are drawn dashed.
One of the challenges is that the \SigCs\ must connect to form a single polygonal curve $\curve$. We therefore ``string'' them together with additional curve pieces, which we refer to as {\em \SCurves}; these are drawn in purple in the figures. In order to ensure that $\curve$ admits a simple embedding, we construct the gadgets (and in particular the different curve pieces) in the order defined by a specific depth-first traversal of $G_F$.

This traversal starts in an arbitrary variable-vertex and explores $G_F$ in depth-first fashion; adjacent vertices are visited counter-clockwise.  This yields a simple traversal curve, which traces around the depth-first tree. 
We refer to the resulting order that vertices and edges are visited as the \emph{\SOrder} on $G_F$. In particular, every variable vertex $v$ has a unique clause vertex from which it is discovered first; we refer to this as the {\em previous clause} of $v$.
For technical reasons, our construction only allows backtracking in a clause-vertex. Therefore, if the traversal considers traversing an edge $e$ from vertex $v$ to clause $c$, then $e$ is always explored and if the $c$ has been visited before then the traversal backtracks in $c$ and continues back across $e$. If the traversal considers traversing an edge $e$ from a clause $c$ to a vertex $v$, then $e$ as well as $v$ are only (re-)visited if $v$ is discovered for the first time, i.e., if $c$ is the previous clause of $v$. Otherwise, the traversal backtracks in $c$ directly and does not explore the edge $e$ to $v$. 

We construct the gadgets in the \SOrder.
%
For clarity of exposition, we first provide a more condensed description of the curve pieces within each gadget: Figures \ref{FDGadget-PartialSync-01}, \ref{FDGadget-Sync-01v2}, \ref{FDGadget-Clause-01}a-b) describe \SigCs\ and \SCurves\ that connect in a tree structure, with some curve pieces having {\em terminals}. In Section \ref{Polyholes-Linking} and Figure \ref{FDGadget-Linking-01} we describe how to convert the tree into the simple curve $\curve$ by essentially performing a planar tree traversal of the tree. 



\subsection{WIRE Gadget \label{Polyholes-WireGadget}}
%
\begin{wrapfigure}{R}{0.26\textwidth}
\vspace*{-6ex}
\includegraphics[width=0.24\textwidth]{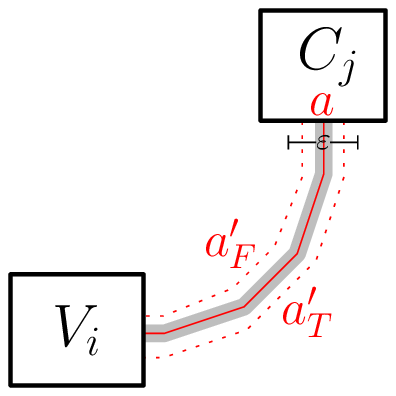}
\vspace*{-3ex}
\end{wrapfigure}
 Our WIRE consists of a hole in $\surface$ and a polygonal \SigC\
 which runs from the variable to the clause.  
The curve is
 positioned such that a valid embedding can be on either side of the hole.
Which side the embedded curve lies on
 determines the value of the variable, or in other words
the {\em signal} of the wire.
 Mapping a WIRE on the right side of the hole constitutes a TRUE
 assignment while mapping a WIRE on the left is a FALSE assignment,
 where the direction of the WIRE is determined by the \SOrder.
 Clearly, any embedded curve cannot
 change which side of the hole it lies in since it needs to be a
 continous curve in $\surface$. The complexity of each WIRE is constant.

All our remaining gadgets contain collections of pairs of
holes and signal curves, where the sides of the hole encode the same TRUE/FALSE assignments as in a WIRE. We therefore also consider \SigCs\ to {\em carry a signal}. Unlike in a regular WIRE, those gadgets will impose additional constraints on the signal curves in order to implement the desired functionality of the gadget.


\subsection{BASE Gadget \label{Polyholes-SubGadget}}
\begin{figure*}[t]
\centering \includegraphics[width=0.99\textwidth]{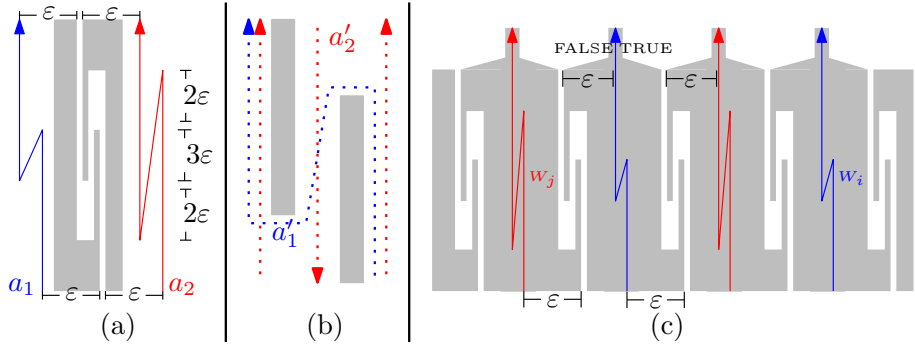}
    \caption{  a) The BASE gadget is used to constrain the mapping of pairs of WIREs.  $a_1$ and $a_2$ are polygonal curves and the gray regions are holes.
		b) Curves $a_1'$ and $a_2'$ are forced to intersect.
		c) Multiple copies can be placed in parallel to propagate constraints.
    \label{FDGadget-PartialSync-01}}
\end{figure*}

The BASE gadget serves as a subgadget in the VARIABLE, NOT, and CLAUSE gadgets, where it is used to constrain the values carried on a pair of WIREs.  The BASE gadget is formed by a pair of holes and two polygonal \SigCs\ $a_1$ and $a_2$, see Figure \ref{FDGadget-PartialSync-01}(a).  Both $a_1$ and $a_2$ cannot be mapped within \FD\ $\varepsilon$ to any embedded curves between the holes because they would cross, see Figure \ref{FDGadget-PartialSync-01}(b).  Note that even if the width of the BASE gadget is reduced arbitrarily the mapped curves would still cross.  
The complexity of a single base gadget is constant.
By having multiple signal curves passing through multiple copies of parallel BASE gadgets, we can propagate constraints, see Figure \ref{FDGadget-PartialSync-01}(c).  In the figure, if in a valid embedding the  signal curve $W_{i}$ is mapped to its FALSE side then all subsequent \SigCs\ to the left must map to their FALSE side.  Likewise, if in a valid embedding the \SigC\ $W_{j}$ is mapped to its TRUE side then all subsequent \SigCs\ to the right must map to their TRUE side.

\subsection{VARIABLE Gadget \label{Polyholes-SyncGadget}}
\begin{figure*}[t]
\centering \includegraphics[width=0.99\textwidth]{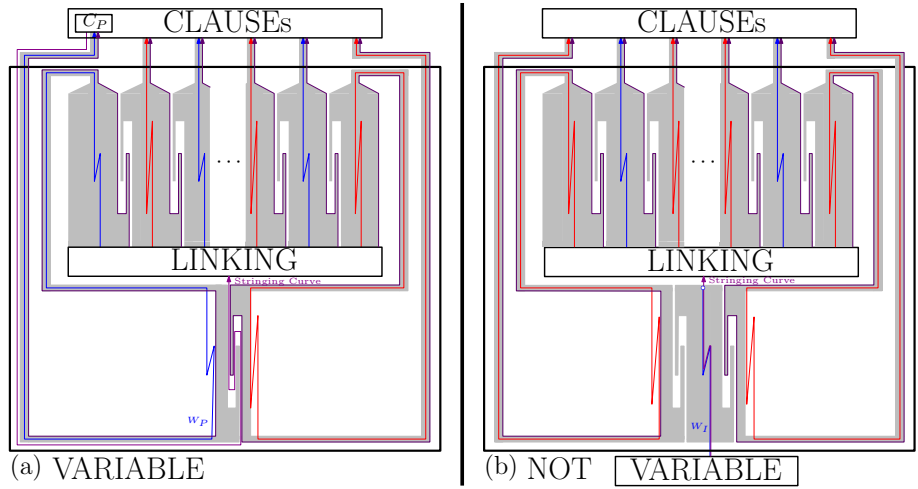}
    \caption{  a) The VARIABLE gadget ensures all of the copies of a variable carry the same signal. 
		b) The NOT gadget takes a variable WIRE as input and creates multiple WIREs carrying the opposite signal.
    \label{FDGadget-Sync-01v2}}
\end{figure*}

The VARIABLE gadget is used to ensure all wires associated with a single variable carry the same value. 
 At its core, the VARIABLE gadget is multiple BASE gadgets in parallel.  As shown in Section \ref{Polyholes-SubGadget}, parallel BASE gadgets can be used to propagate constraints between \SigCs.  Since, we want all of our WIREs to carry the same signal we need to construct a ring of BASE gadgets, see Figure \ref{FDGadget-Sync-01v2}(a).  The BASE gadget at the bottom of the VARIABLE gadget is used to compare the two \SigCs\ on the ends of the parallel BASE gadgets to form a ring. 
 From the properties of the BASE gadget follows that in a valid embedding all of the \SigCs, and therefore all of the WIREs exiting the VARIABLE, carry the same value.  The complexity of the VARIABLE gadget is linearly dependent on the degree of the vertex it represents in $G_F$. After being synchronized, the WIREs then continue to their respective CLAUSE gadgets.

Let $C_P$ be the previous CLAUSE of this VARIABLE (as defined by the \SOrder, see Section \ref{Polyholes-StringingOrder}), and let $W_P$ be the WIRE that connects $C_P$ to this VARIABLE.  The \SCurve\ is constructed such that it follows the FALSE side of the hole of $W_P$ from the CLAUSE to the VARIABLE gadget, see Figure \ref{FDGadget-Sync-01v2}(a).  The \SCurve\ enters the VARIABLE gadget through the bottom BASE gadget.  The portion of the \SCurve\ which passes through the BASE gadget has a shape between that of the two curves which may pass through the BASE gadget.  This allows it to map to an embedded curve which passes through the BASE gadget without crossing any embeddings of the two signal curves of the base gadget.

Note that we can assume without loss of generality that $C_P$ and the VARIABLE are directly connected via WIRE $W_P$, without any NOT gadget in-between. If the variable appears negated in $C_P$, we can change all instances of this variable from positive to negative and vice versa in $F$ without changing the satisfiability of the formula.

\subsection{NOT Gadget \label{Polyholes-NotGadget}}

The NOT gadget takes a WIRE $W_I$ as input and creates multiple WIREs which have the signal $\overline{W_I}$.  The structure of the NOT gadget is similar to the VARIABLE gadget.  The key difference is that the NOT gadget takes a WIRE, $W_I$ from a VARIABLE gadget as input.  Once again the gadget is constructed using a ring of BASE gadgets.  However, $W_I$ is directed opposite to the \SigCs\ of the NOT gadget.  Thus, if in a valid embedding $W_I$ is mapped to its TRUE side it will force all of the \SigCs\ from the NOT gadget to map to their FALSE side.  Note that we actually need only one of these \SigCs/WIREs to connect to a CLAUSE gadget.  The other WIREs terminate after leaving their associated BASE gadget.  Thus the complexity of the NOT gadget is constant.

Similar to the VARIABLE gadget, a \SCurve\ links this NOT gadget to the previous gadget in the \SOrder. Here, the \SCurve\ passes through the same BASE gadget as $W_I$ by following the path of $W_I$, which in turn allows the \SigCs\ to be mapped such that they do not cross.


\subsection{LINKING Gadget \label{Polyholes-Linking}}
\begin{figure*}[t]
\centering \includegraphics[width=0.99\textwidth]{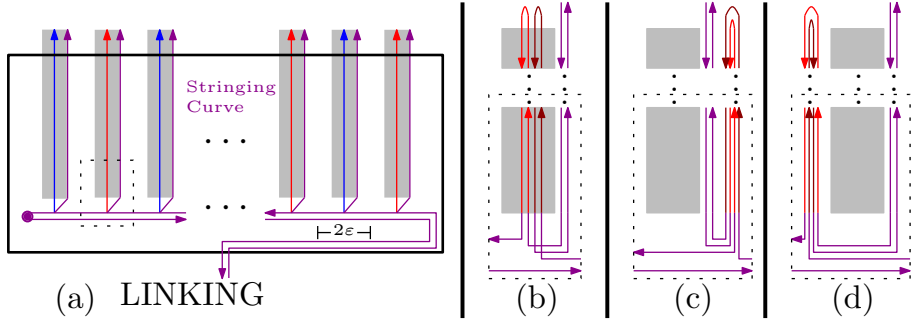}
    \caption{   a) The WIREs are initialized in the LINKING gadget.
		b) This shows the topology of the curves associated with this WIRE.
		c,d) There exists a valid embedding regardless if the curve is on the TRUE or FALSE side of the WIRE.
    \label{FDGadget-Linking-01}}
\end{figure*}

%
%
The LINKING gadget connects the \SigCs\ of the VARIABLE gadget (or NOT gadget) with a \SCurve\ in order to form the compact representation of a tree, see Figure \ref{FDGadget-Linking-01}(a). There exists a valid embedding of this \SCurve\ regardless of the values of the \SigCs, see Figure \ref{FDGadget-Linking-01}(a).

We now explain how we convert the condensed tree description used in Figures \ref{FDGadget-PartialSync-01}, \ref{FDGadget-Sync-01v2}, \ref{FDGadget-Clause-01}a-b) into the simple curve $\curve$:
In addition to transmitting a signal to a CLAUSE with the WIRE we may also need to travel from the CLAUSE to the next variable in the \SOrder.  For this purpose we have a \SCurve\ which travels with the WIRE to the CLAUSE.   Because the \SCurve\ is only used to connect the gadgets, we construct it such that it always has a valid embedding. In particular, from the LINKING gadget until the CLAUSE we assume that it follows the right hole boundary of its associated WIRE.  To ensure that the curves associated with the WIRE can always be mapped such that they do not cross, we include a second copy of the curve which carries the signal.  The \SCurve\ is connected between the two copies of the \SigC, see Figure \ref{FDGadget-Linking-01}(b).


The complexity of the LINKING gadget depends linearly on the complexity of the degree of the vertex it is associated with.

\subsection{CLAUSE Gadget \label{Polyholes-ClauseGadget}}

\begin{figure*}[pb]
\centering \includegraphics[width=0.900\textwidth]{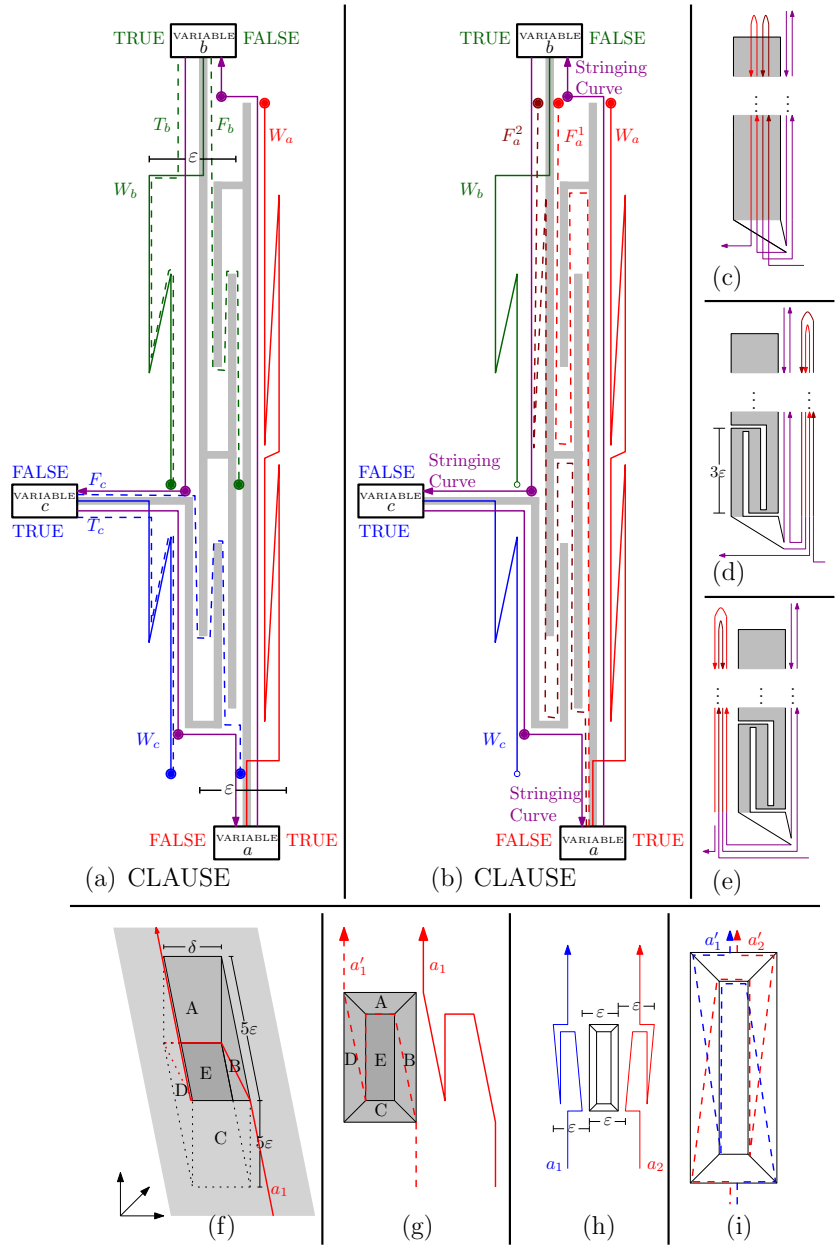}
    \caption{
 a-b) The CLAUSE gadget ensures at least one of its input WIREs has to be mapped to its TRUE side. Valid embeddings are drawn dashed.  
                     c) This shows the topology of the curves including the \HCurve.
		d-e) The curves can still be mapped to the TRUE or FALSE side of the WIRE such that they are simple.  
		f) A deep valley in the surface limits the possible valid embeddings of a curve $a_1$ is a curve.
		g) Valleys and valley curves are represented as shown.
		h) A BASE gadget is formed by a valley and two valley curves.
		i) The embedded curves which lie on faces B and D will cross. \label{FDGadget-Clause-01} \label{FDGadgetPlane-Linking-01} \label{WFDGadget-PartialSyncMulti-01} }
\end{figure*}

This gadget takes three WIREs $W_{v_a}$, $W_{v_b}$, and $W_{v_c}$, as input which correspond to the variables in the clause.  The gadget consists of two BASE gadgets which are slightly narrower, see Figure \ref{FDGadget-Clause-01}.  
  The gadget ensures that if all three variables are false then no valid embedding exists because any embedding of the curves will cross.  Likewise, if any of the variables are true then there exists a valid embedding of the curves.

The \SigCs\ $W_{v_b}$ and $W_{v_c}$ can each map to one of the BASE gadgets.  In particular, if valid embeddings for $W_{v_b}$ and $W_{v_c}$ lie on the FALSE side of their hole then they must pass through the BASE gadget, see Figure \ref{FDGadget-Clause-01}(a).  For a valid embedding on the TRUE side there is no restriction. The \SigC\ $W_{v_a}$ can map to both of the BASE gadgets.  In particular, if a valid embedding for $W_{v_a}$ lies on the FALSE side then it must pass through at least one of the two BASE gadgets, see Figure \ref{FDGadget-Clause-01}(b).  


If $W_{v_a}$ is mapped to $F^1_{v_a}$ then $W_{v_b}$ cannot be mapped to $F_{v_b}$ since the BASE gadget would cause the curves to cross.  Thus, in this case, any valid embedding for $W_{v_b}$ must lie on the TRUE side of its hole.  Likewise, if $W_{v_a}$ is mapped to $F^2_{v_a}$ then $W_{v_c}$ cannot be mapped to $F_{v_c}$ since the BASE gadget would cause the curves to cross.  Thus, in this case, any valid embedding for $W_{v_c}$ must lie on the TRUE side of its hole.  Finally, note that $W_{v_b}$ may be  mapped to $F_{v_b}$ and $W_{v_c}$ mapped to $F_{v_c}$ without crossing.   $W_{v_a}$ cannot pass through either of the BASE gadgets and any valid embedding must lie on the true side of its hole.

From this we see that the \SigCs\ associated with a CLAUSE gadget have a valid embedding if and only if at least one of the \SigCs\ is mapped to the TRUE side of its hole.  As seen in Figure \ref{FDGadget-Clause-01}(a), we must also consider how the \SCurve\ travels through the clause from one variable to another in the CLAUSE gadget when we construct the portion of the CLAUSE gadget associated with a particular WIRE.  If the VARIABLE gadget associated with the WIRE counterclockwise from the current one has not been visited yet then we trace the \SCurve\ along the FALSE side of the not yet constructed WIRE.  We then construct the VARIABLE as described in Section \ref{Polyholes-SyncGadget}.  Otherwise, we return to the previous gadget on the current WIRE following along the \SCurve\ as explained in Section \ref{Polyholes-Linking}.  Since a CLAUSE gadget always takes three WIREs as input it has constant complexity.

\subsection{ Main Theorem \label{Polyholes-Proof}}

The reduction outlined in this section takes polynomial time.  In particular, the gadgets can be constructed in the \SOrder.  Each gadget has either polynomial or constant complexity and the number of gadgets required is directly dependent on the complexity of the planar graph. 
 In addition, we have seen how a satisfying assignment of the variables in the 3SAT formula is possible if and only if there exists a valid embedding, i.e., a simple curve within \FD\ $\varepsilon$ of the target curve.  
This yields Theorem \ref{Polyholes-Thm}.  

%

\section{Extensions to the Curve Embedding Problem \label{Extensions}}


\subsection{NP-Hardness in the Plane \label{Plane-Proof}}

Surprisingly, we can show that the \problem\ is NP-hard even when the surface is just a plane.  Once again we reduce from solving planar 3SAT.  The key insight is that we can simulate the holes in our original reduction by using curves which we call \emph{\HCurves}.  If we trace a curve $\varepsilon$ distance above every hole boundary, the curve it maps to will act as a similar boundary to the WIRE gadgets as the original hole.  Like the \SCurve, \HCurves\ are formed in the LINKING gadgets, see Figure \ref{FDGadgetPlane-Linking-01}(c).

On the way to the CLAUSE the \HCurve\ follows the TRUE side of the hole and then returns on the FALSE side of the hole.  The \SigCs\ of LINKING gadgets can still be mapped to simple curves regardless of the value of the WIRE, see Figure \ref{FDGadgetPlane-Linking-01}(d)(e).  Using this method all of the holes of our original reduction can be simulated using \HCurves\ since every hole is associated with a WIRE.  One minor issue is that we need to ensure that the curves associated with the WIRE cannot be mapped inside the hole.  To prevent this we add a zigzag to the \HCurve, see Figure \ref{FDGadgetPlane-Linking-01}(d)(e).  Because the zigzag is $3\varepsilon$ long none of the WIRE's curves can be mapped inside the simulated hole.
%

\begin{theorem}%
    \label{PlaneEmbedding-Thm}
The decision version of the \SCEP\ is NP-hard under \FD\ for a polygonal curve and a plane.
\end{theorem}

\subsection{NP-Hardness for the Weak Fr\'{e}chet Distance on Terrains  \label{WFD-Proof}}

We now consider the \problem\ using the \WFD\ as the distance measure.  The reduction given in Section \ref{Polyholes} cannot be applied directly.  In particular, it is easy to show that the BASE gadget no longer forces the curves to cross if the \WFD\ is used.  We reuse much of the original reduction from planar 3SAT except for a new BASE gadget which works for \WFD, see Figure \ref{WFDGadget-PartialSyncMulti-01}(f)(g)(h)(i).  The gadget uses a deep valley rather than monotonicity to restrict the valid embeddings of the curves.  Using this we prove that the \problem\ is NP-hard under \WFD\ when the surface is a 3d terrain.  See Appendix \ref{WFD-Reduction} for details regarding the gadgets.


\begin{theorem}%
    \label{WFD-Thm}
The \problem\ is NP-hard to compute under \WFD\ for a polygonal curve and a 3d terrain.
\end{theorem}

\subsection{Partial Fr\'{e}chet Distance between a Polygonal Strip and a Plane\label{PartialFD-Proof}}

We reduce from the \problem\ under \FD\ for a curve and a plane, see Section \ref{Plane-Proof}. We treat the curve as a very narrow surface, as mentioned in Section \ref{PolyTime}. Since this surface is based on a 3d polygonal curve, it can be trivially triangulated such that its dual graph is a path, and hence the surface is a polygonal strip.

\begin{theorem}%
    \label{PartialFD-Thm}
The partial \FD\ is NP-hard to compute between a polygonal strip and a plane.
\end{theorem}

\section{Conclusions \label{Conclusions}}

We introduced the \problem, studied polynomial-time solutions for some variants, and proved NP-hardness in $\R^3$ under \FD\ and the \WFD. It appears that hardness is caused by the simplicity requirement. 
We conjecture that the problem is also NP-hard under \FD\ and the \WFD\ if (1) $\surface$ is a plane and $\curve$ is coplanar and self-intersecting, and if (2) $\surface$ is a plane with disconnected holes and $\curve$ is coplanar and simple.

\subsubsection{Acknowledgements \label{Acknowledgements}}

The authors thank Sariel Har-Peled, Anne Driemel, and Benjamin Raichel for initial discussion of the \problem.


\bibliographystyle{plain}%
\bibliography{CurveNPhard}%

\newpage
\begin{appendix}

\section{Weak Fr\'{e}chet Distance Reduction \label{WFD-Reduction}}

In this section we outline a reduction from planar 3SAT to the \problem\ under \WFD\ for 

\subsection{WIRE Gadget \label{WFD-WireGadget}}

The WIRE gadget described in Section \ref{Polyholes-Linking} will still work for the \WFD\ reduction with one minor change.  Rather than holes we can use deep valleys in the surface.  As before a valid embedding of the curve must be continuous so it cannot cross the deep valley.

\subsection{BASE Gadget \label{WFD-SubGadget}}

We construct a completely different BASE gadget from the one presented in Section \ref{WFD-SubGadget}.  Our previous gadget used the monotonicity of the \FD\ to restrict the mappings of a pair of curves.  For \WFD\ we used deep valleys in the surface to restrict the mappings of a pair of curves,  see Figure \ref{WFDGadget-PartialSyncMulti-01}(f).  Let the width of the valley, $\delta$, be strictly less than $\varepsilon/4$ in our construction.  The curve $a_1$ must cross both face B and D of the valley,  see Figure \ref{WFDGadget-PartialSyncMulti-01}(g).  A pair of curves cannot both be embedded in a single BASE gadget,  see Figure \ref{WFDGadget-PartialSyncMulti-01}(h)(i).

\begin{figure*}[t]
\centering \includegraphics[width=0.99\textwidth]{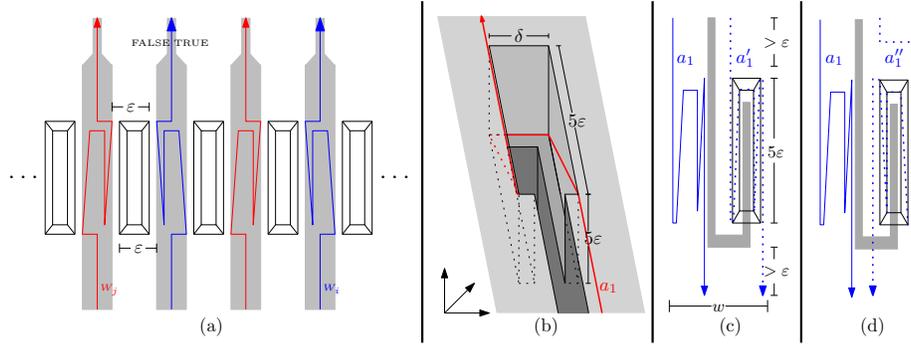}
    \caption{  a) The BASE gadgets can be placed in parallel as before.
		b)  For the CLAUSE gadget we add a deeper valley in the original valley.
		c)  $a'_1$ is a valid embedding of the curve $a_1$.
		d)  $a_1$ cannot be embedded as a curve which crosses the faces of BASE gadget in the opposite order.
    \label{WFDGadget-PartialSyncMulti-02}}
\end{figure*}

We can thus use this BASE gadget to form our other gadgets with one exception.  The previous CLAUSE gadget placed the \SigCs\ close to the BASE gadget.  In this new BASE gadget, the \SigC\ being close to the valley allows it to map to the faces in the opposite order.  By adding a deeper valley to the BASE gadget we can restrict the order the embedded curve crosses the faces, see Figure \ref{WFDGadget-PartialSyncMulti-02}(a)(c)(d).

\subsection{VARIABLE Gadget \label{WFD-SyncGadget}}

The VARIABLE gadget is constructed with a ring of BASE gadgets similar to the one in Section \ref{Polyholes-SyncGadget}, see Figure \ref{WFDGadget-Sync-01}(a).  Note that the \SCurve\ from the previous CLAUSE in the \SOrder\ can be mapped on either side of its associated valley in the bottom BASE gadget.  This ensures the \SCurve\ has a valid embedding regardless of whether the VARIABLE is assigned the value TRUE or FALSE.

\subsection{NOT Gadget \label{WFD-NotGadget}}

\begin{figure*}[t]
\centering \includegraphics[width=0.99\textwidth]{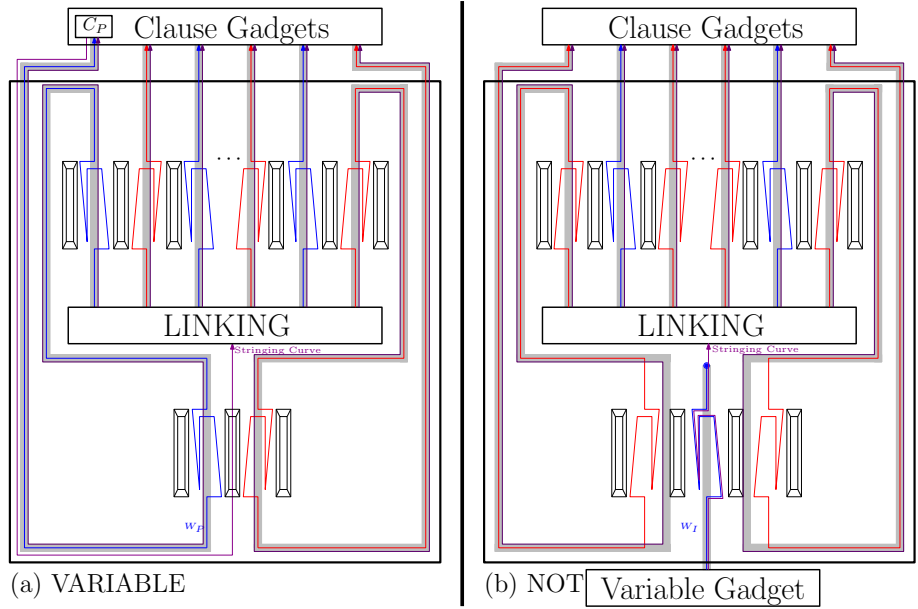}
    \caption{   a) The VARIABLE gadget ensures all of the copies of a variable carry the same signal. 
		b) The NOT gadget takes a variable WIRE as input and creates multiple WIREs carrying the opposite signal.
    \label{WFDGadget-Sync-01}}
\end{figure*}

The NOT gadget is constructed with a ring of BASE gadgets similar to the VARIABLE gadget in Section \ref{WFD-SyncGadget}, see Figure \ref{WFDGadget-Sync-01}(b).  Similar to the NOT gadget from Section \ref{Polyholes-NotGadget}, this gadget takes a WIRE, $W_I$, as input and, since the input WIRE is directed in the opposite direction the NOT gadget will be assigned to $\overline{W_I}$.  As with the original NOT gadget, the \SCurve\ follows the signal curve of the input WIRE $W_I$ and, thus, there exists a valid embedding of the \SCurve\ by mapping it to the same BASE gadget as the signal curve.  

\subsection{LINKING Gadget \label{WFD-Linking}}

The LINKING gadget described in Section \ref{Polyholes-Linking} will still work for the \WFD\ reduction with the hole replace with a deep valley.

\subsection{CLAUSE Gadget \label{WFD-ClauseGadget}}

\begin{figure*}[t]
\centering \includegraphics[width=0.99\textwidth]{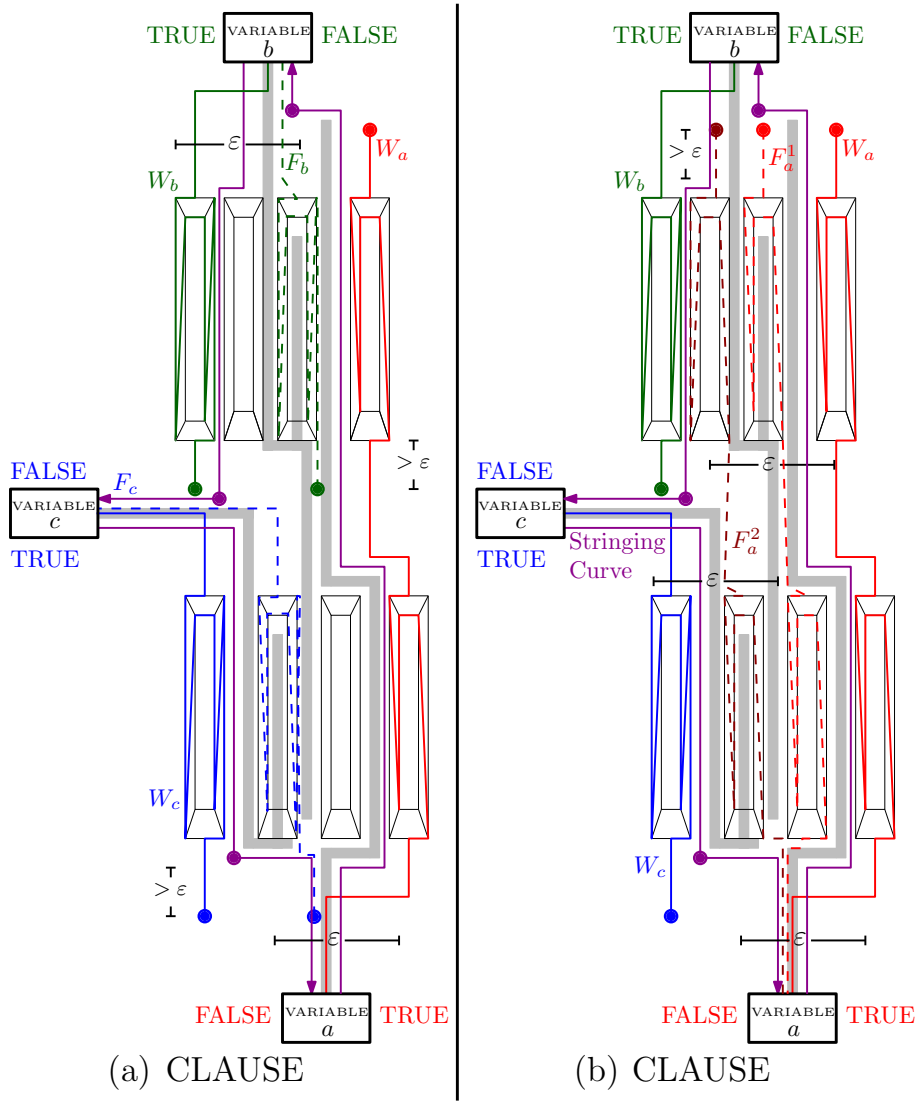}
    \caption{    a-b) The CLAUSE gadget ensures at least one of its input WIREs must be mapped to its TRUE side. Valid embeddings are drawn dashed.  
    \label{WFDGadget-Clause-01}}
\end{figure*}

The CLAUSE gadget is constructed to from several BASE gadgets, similar to the one in Section \ref{Polyholes-ClauseGadget}.  As before it takes three WIREs as input and admits a valid embedding of the signal curves if and only if at least one of them is mapped to the TRUE side of its WIRE, see Figure\ref{WFDGadget-Clause-01} .  As mentioned in Section \ref{WFD-SubGadget}, we use a variant of the BASE gadget which allows the signal curve close to the valley of the BASE gadget.  

The \SigCs\ $W_{v_b}$ and $W_{v_c}$ can each map to one of the BASE gadgets.  In particular, if valid embeddings for $W_{v_b}$ and $W_{v_c}$ lie on the FALSE side of their hole then they must pass through the deep valley BASE gadget, see Figure\ref{WFDGadget-Clause-01}(a).  For a valid embedding on the TRUE side there is no restriction. The \SigC\ $W_{v_a}$ can map to both of the deep valley BASE gadgets.  In particular, if a valid embedding for $W_{v_a}$ lies on the FALSE side then it must pass through at least one of the two BASE gadgets, see Figure \ref{FDGadget-Clause-01}(b).  The proof of correctness follows from the same argument as for the CLAUSE gadget in Section \label{Polyholes-ClauseGadget}.

\end{appendix}

\end{document}